\documentclass[%
 reprint,
 amsmath,amssymb,
 aps,
]{revtex4-2}

\usepackage{graphicx}
\usepackage{dcolumn}
\usepackage{bm}
\usepackage{braket}

\begin{document}

\preprint{APS/123-QED}

\title{Determination of principal axes orientation in an ion trap using matter-wave interference}

\author{Ryoichi Saito$^{1}$}
\email{r-saito@phys.titech.ac.jp}
\author{Takashi Mukaiyama$^{1}$}%
\email{mukaiyama@phys.titech.ac.jp}
\affiliation{
$^1$Department of Physics, Tokyo Institute of Technology, Ookayama 2-12-1, Meguro-ku, Tokyo 152-8550, Japan\\
}%

\date{\today}

\begin{abstract}
We investigate the control mechanism of trap frequencies and determine the orientation of ion trap principal axes. The application of DC voltage to the ground electrodes, commonly employed to finely tune trap frequencies in ion traps, leads to the rotation of the trap principal axes. Analyzing the ion matter-wave interference signal enables us to determine the directions of the trap principal axes. Both the experiments and simulations reveal an avoided-crossing behavior resulting from the coupling between the trap radial axes. Additionally, simulations indicate that symmetric trap structures lack this coupling, suggesting that trap structure asymmetry causes coupling between the axes. The findings of this study offer valuable insights into ion traps for diverse applications in quantum science and technology. 
\end{abstract}

\maketitle


\section{\label{sec:level1}Introduction}
Paul traps, which are used for trapping and guiding charged particles, have diverse applications across multiple disciplines, including mass spectroscopy, plasma physics, atomic and molecular physics, and quantum physics~\cite{R C Thompson_1990}. Notably, the development of ion traps as quantum platforms that rely on trapping and cooling atomic ions constitutes a significant advancement in quantum computing~\cite{HAFFNER2008155}, quantum measurements~\cite{PhysRevLett.116.063001, 10.1126/science.1114375}, and sensing applications~\cite{Biercuk2010, aao4453, Shaniv2017, 10.1063/5.0046121, PhysRevApplied.16.044007, PhysRevX.7.031050, Noguchi2014,Campbell_2017}. These fields have experienced considerable growth in recent years.

The spatial confinement of ions in a Paul trap results from a combination of radio-frequency (RF) and DC electric fields, which effectively trap ions within a pseudoharmonic potential. The design of the ion trap correlates closely with the shape of the trap potential.
Following the initial conception of the original quadrupole ion trap, the trap structure has undergone subsequent advancements to enhance its optical accessibility,
exemplified as an end-cap trap~\cite{SCHRAMA199332} configuration, which demonstrates superior capabilities for trapping single ions and is commonly employed in the development of ion-trap optical frequency standards.
Stylus traps are expected to be useful in applications such as surface property sensing owing to their accessible structure~\cite{Maiwald2009, 10.1063/1.4817304}.
Additionally, planar or microfabricated traps offer the advantage of constructing diverse trap geometries owing to their high degrees of freedom in electrode shaping on a substrate~\cite{PhysRevLett.96.253003, Amini_2010}.In contrast, linear traps~\cite{Nagerl1998, Schmidt-Kaler2003} are an excellent platform for purposes such as quantum information processing and quantum computing~\cite{PRXQuantum.2.020343}.
These traps excel in trapping one-dimensional ion chains, with radial confinement achieved through RF electric fields and axial confinement maintained by DC electric fields.

The direction of the trap axis corresponds to the principal axis of ion motion and is a crucial factor in controlling ion motion. The trap axis is determined mainly by the structure of the ion-trap electrodes. From the perspective of efficient laser cooling, asymmetric trap electrodes are widely utilized to lift the degeneracy of trap frequencies. In such cases, the direction of the trap axes becomes nontrivial from the structure itself. Moreover, a DC voltage is occasionally applied to the DC electrodes to finely adjust the trap frequencies in an ion trap. The application of this DC voltage distorts the trap potential, causing rotation of the principal axes of the trap. This phenomenon poses a challenge when precise control over the direction of ion motion is required. Nevertheless, there is no experimental instance that clearly shows the determination of the trap principal axis directions, and no meticulous investigation has been conducted on the rotation of trap axes resulting from the application of a DC voltage to DC electrodes.

In this study, we successfully determined the orientation of the principal axes of an ion trap by analyzing the matter-wave interference signal of an ion. To achieve this end, we applied a Raman transition to the ion, imparting a momentum kick along the propagation direction of the Raman laser~\cite{PhysRevLett.126.153604}. This process excites the ion motion along laser propagation direction, and the ion motion is decomposed into the motions along the principal axes of the ion trap. The matter-wave interference signal exhibits a complex peak structure that encapsulates the details of the angles between the direction of the momentum kick and the trap principal axes. By interpreting this signal, the directions of the principal axes can be precisely determined. Notably, when applying a DC voltage to the ground electrodes, we observe not only a shift in the trap frequency but also a rotation of the trap principal axes. To further validate these findings, we conduct electromagnetic field simulations, providing a theoretical confirmation of the rotational behavior of the trap axes along with the trap frequency changes.

\section{\label{sec:level2}Experimental apparatus and procedure}
\begin{figure}[tb]
	\centering
	\includegraphics[width=8.5cm]{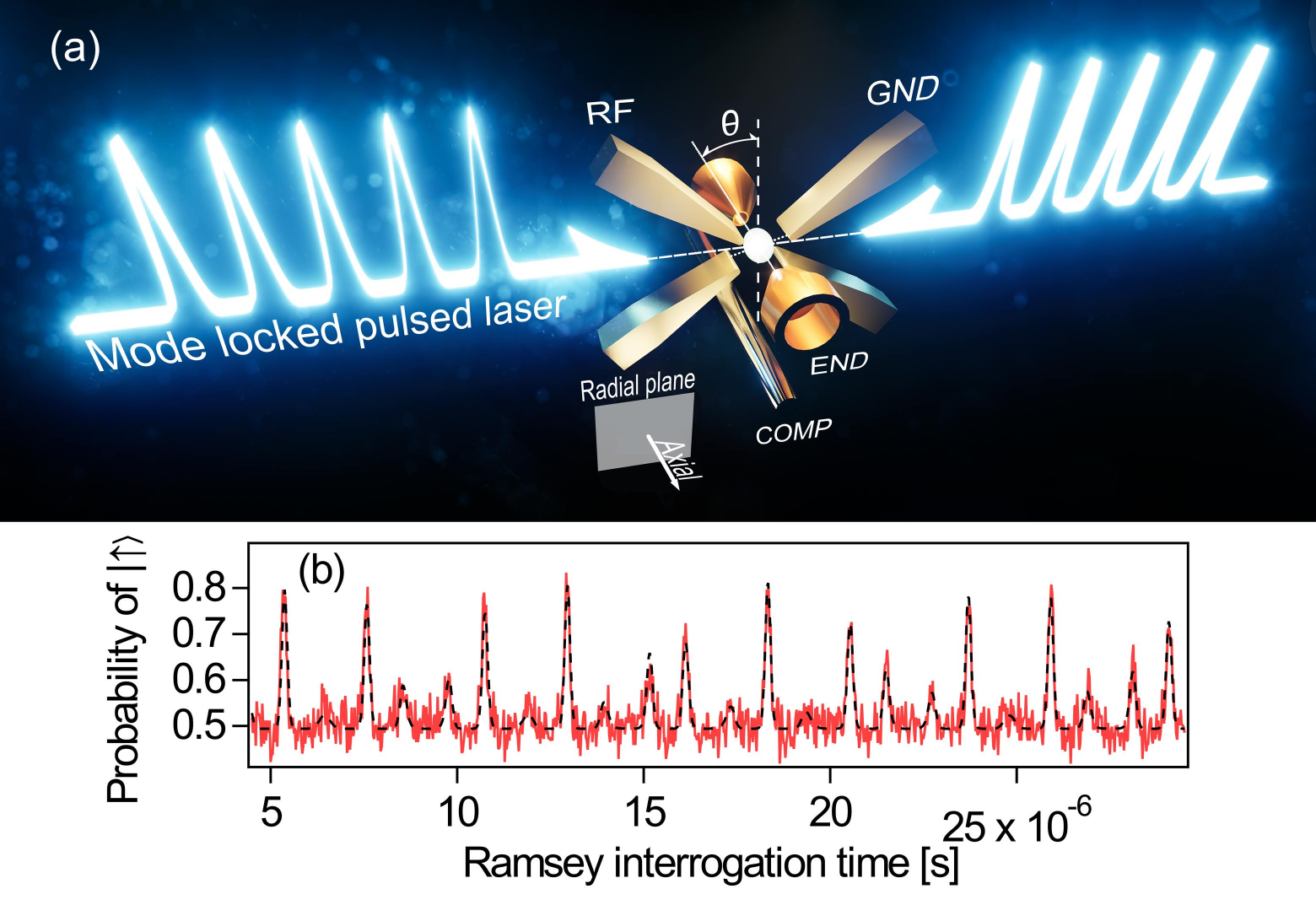}
	\caption{
	(a) Schematic image of the experimental apparatus of an ion matter-wave interferometer.
	We confine a single $^{171}\rm Yb^+$ ion within a conventional linear Paul trap.
	The $\rm Yb^+$ ion is subjected to Doppler cooling by laser beams incident from three directions.
	The wavelengths of the cooling, repumping 1, and repumping 2 lasers are $369$, $935$, and $760~\rm nm$, respectively.
	The mode-locked pulsed laser is applied oppositely in radial directions of the trap to kick the ion.
	(b) A typical matter-wave interference signal obtained in the experiment.
	}
	\label{fig:fig1}
\end{figure}

We control the trap confinement of a $^{171}\rm Yb^+$ ion in a conventional linear ion trap as shown in Fig.~\ref{fig:fig1}(a). The trap electrode configuration consists of two RF electrodes, two ground (GND) electrodes, two END electrodes, and one COMP electrode. The RF and GND electrodes are thin blade-shaped components, whereas the END electrodes are circularly truncated cones with central apertures. The tips of the RF and GND electrodes are arranged to form a rectangle with an aspect ratio of approximately 5:7.  Typically, we apply an RF voltage with a frequency of $24~\rm MHz$ and amplitude of $300~\rm V$ to the RF electrodes with the GND electrodes functioning as the RF ground. The END electrodes are supplied with voltages ranging from approximately $3$ to $5~\rm V$. These voltage settings result in trap confinement of approximately $1~\rm MHz$ along the three axes. Micromotion is compensated by applying a voltage to one of the GND electrodes, and a rod-shaped COMP electrode is placed $2.7~\rm mm$ away from the trap center~\cite{10.1063/5.0046121}.

We apply a DC voltage to both GND electrodes to regulate both the trap frequency and orientation of the trap axis relative to the electrodes. While applying the same DC voltage to both GND electrodes for overall confinement control, one of the GND electrodes receives an additional voltage to compensate for micromotion. This compensation voltage is integrated with the confinement control voltage. The trap frequencies, controlled by the DC voltage applied to the GND electrode, are measured using the resonant modulation scheme commonly known as the "tickle method," or fast displacement method~\cite{PhysRevA.104.053114, 10.1063/5.0100007}.

A laser at 369~nm is applied to cool the ion and two additional lasers at $935~\rm nm$ and $760~\rm nm$ are applied for repumping purposes.
We cool the trapped $^{171}\rm Yb^+$ ion using conventional Doppler cooling, and the ion fluorescence is collected using an objective lens and transferred to an EMCCD camera and a photomultiplier tube (PMT) located above the ion trap.
The PMT signal is used to determine the population of a $^{171}\rm Yb^+$ ion in the qubit states $\Ket{^2S_{1/2}, F=0, m_F=0}=\Ket{\downarrow}$ and $\Ket{^2S_{1/2}, F=1, m_F=0}=\Ket{\uparrow}$.

A mode-locked pulsed laser with a wavelength of $355~\rm nm$ is applied in a counterpropagating configuration from the radial direction of the trap, as illustrated in Fig.~\ref{fig:fig1}(a), to observe ion matter-wave interference. This pulsed laser drives the Raman transition between the two qubit states, $\Ket{\uparrow}$ and $\Ket{\downarrow}$.
To initiate and close the interferometer, a $\pi/2$ pulse composed of 10 laser pulses is applied to the ion. The duration of the $\pi/2$ pulse is approximately 80 ns, which is approximately $1/10$ the trap period. Initially prepared in the $\Ket{\downarrow}$ state, the applied $\pi/2$ pulse transfers half of the ion population to the $\Ket{\uparrow}$ state and imparts a momentum kick of $2 \hbar k$, where $k$ represents the wavenumber of the laser. Consequently, this process generates the spin-motion entanglement of ion~\cite{PhysRevLett.105.090502, PhysRevLett.104.140501, PhysRevLett.110.203001}.

In our interference experiments, the first $\pi/2$ pulse is applied to an ion in the $\Ket{\downarrow}$ state. Subsequently, half the population in $\Ket{\uparrow}$ receives a kick and the ion wave packet oscillates in the trap potential. Following the free-evolution time, the second $\pi/2$ pulse is irradiated to close the interference. Finally, the ion is irradiated with a $369~\rm nm$ laser to detect the fluorescence and determine its internal state.
Figure~\ref{fig:fig1}(b) shows a typical interference signal obtained in the experiment. The vertical axis represents the probability of finding the ion in the $\Ket{\uparrow}$ state, with the peak height indicating the extent of the overlap between the separated ion wave packets at each interrogation time. The interference signal reflects the relative angle between the laser propagation direction and the trap principal axes. The direction of the trap principal axes can be determined by fitting the interference signal to a theoretical curve.

\section{\label{sec:level3}Determination of the orientation of the trap principal axes}

\begin{figure*}[tb]
	\centering
	\includegraphics[width=16cm]{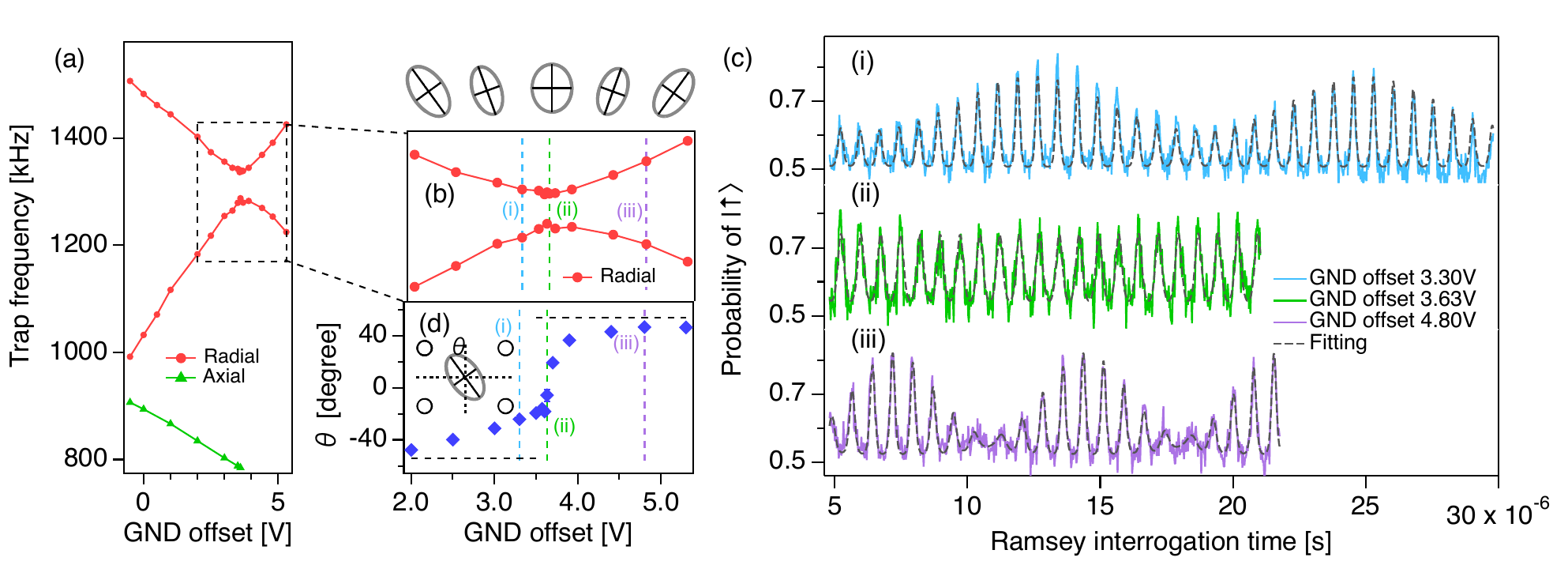}
	\caption{
	(a) Experimentally measured trap frequencies as a function of the GND offset voltage. (b) The expanded plot of the result around the coupling point.
	(c) Experimentally obtained interference signals at GND offset voltages of 3.30, 3.63, and 4.80~V, with the fitting results shown with the dashed curves.
	(d) The angle $\theta$ determined from the fitting as a function of the GND offset voltage. The rotation of the trap principal axes around the coupling point is seen.}
	\label{fig:fig2}
\end{figure*}

Figures~\ref{fig:fig2}(a) and (b) illustrate the measured trap frequency with the application of the GND offset voltage. The filled triangles in Fig.~\ref{fig:fig2}(a) show the measured trap frequencies in the axial direction along the line connecting the two END electrodes. The filled circles in Fig.~\ref{fig:fig2}(a) show the measured trap frequencies along the two axes in the radial direction. Initially at 0V GND offset voltage, the trap frequencies of the three axes differ. With an increase in voltage, one axis of the radial confinement strengthens, whereas the other radial axis and axial confinements weaken. Because a positive voltage creates an axial confinement for a positively charged ion, applying an offset voltage to the GND electrodes weakens the axial confinement. 

The two axes in the radial direction exhibit the closest trap frequency at an offset voltage of 3.63~V, but the trap frequencies do not coincide perfectly. This anti-crossing phenomenon indicates coupling between the two radial trap axes. The trap-frequency difference under the closest condition is approximately 50~kHz, indicating the impossibility of matching the trap frequencies of the two radial axes to create a perfectly circular trap potential.

We conduct ion matter-wave interferometry by irradiating pulsed lasers to investigate the rotation of the trap axes around the coupling point of the radial trap frequencies. Figure~\ref{fig:fig2}(c) illustrates the ion matter-wave interference signals. The vertical axis represents the probability of finding an ion in the $\Ket{\uparrow}$ state after the Ramsey sequence, whereas the horizontal axis represents the interrogation time of the Ramsey sequence.
The three interference signals labeled (i), (ii), and (iii) in Fig. ~\ref{fig:fig2}(c) are the results obtained at GND offset voltages of 3.30, 3.63, and 4.80~V, respectively.

Because the pulsed lasers are incident orthogonally to the axial direction of the ion trap, no axial momentum kick is applied to the ion.
Therefore, the ion is expected to undergo two-dimensional motion within the radial plane.
The probability of finding the ion in the $\Ket{\uparrow}$ state, $P$, is expressed as~\cite{PhysRevLett.126.153604}
\begin{widetext}
\begin{equation}
\label{eq:eq1}
P=\frac{1}{2} \left\{ P_{0}+A\exp \left[ - \left( \eta_{1} \sin \theta \right)^2 \left( 2n_{1}+1\right)  (1 - \cos\omega_{1} t ) \right]\exp \left[ -\left( \eta_{2} \cos\theta  \right)^2 \left( 2n_{2}+1 \right) (1 - \cos \omega_{2} t ) \right] \right\}
\end{equation}
\end{widetext}
where $P_0$ is the signal offset, $A$ is the signal amplitude, $\omega_1$ and $\omega_2$ represent the angular trap frequencies of the radial axes, $t$ is the Ramsey interrogation time, and $n_1$ and $n_2$ correspond to the mean photon numbers of each radial axis, formulated as $n = k_{\rm B}T/\hbar\omega$.
$\eta_1$ and $\eta_2$ are the Lamb-Dicke parameters for each radial axis, defined as $\eta = 2k/\sqrt{\hbar/2m\omega}$. In this equation, $\hbar$, $m$ and $k$ represent the Dirac parameter, mass of a $\rm ^{171}Yb^+$ ion, and wavenumber of the mode-lock pulsed laser for the Raman transition, respectively. Thus, $2\hbar k$ is the momentum kick applied to the ion by the Raman transition.
$\theta$ is defined as the angle between the vertical axis and the direction of one of the principal trap axes, as shown in Fig.~\ref{fig:fig1}(a).
The dashed black curves in Fig.\ref{fig:fig2}(c) are the fitting results of the interference signals using Eq.~\ref{eq:eq1}.
Figure~\ref{fig:fig2}(b) shows expanded plots of the experimentally measured radial trap frequencies around the coupling point, indicated by the square in Fig.~\ref{fig:fig2}(a). The vertical dashed lines (i), (ii), and (iii) in Fig.~\ref{fig:fig2}(b) show the GND offset voltages corresponding to the values used to measure the interference signal in Fig.~\ref{fig:fig2}(c).

The interference signals shown in Fig.~\ref{fig:fig2}(c) exhibit peaks when the ion wave packets, separated by the momentum kick, reconverge in the phase space after the interrogation time. Notably, the heights of the peaks gradually change and reveal an envelope, as observed in signals (i) and (iii) in Fig.~\ref{fig:fig2}(c).
This envelope corresponds to the beat frequency, that is, the difference between the two radial trap frequencies.
For signal (i), the peaks of the envelope appear every $\rm 12.5~\mu s$, which is consistent with the frequency difference between the radial trap frequencies of $80.0~\rm kHz$.
Similarly, for signal (iii), the envelope peaks every $\rm 7.25~\mu s$, which is consistent with a frequency difference between the radial trap frequencies of $138~\rm kHz$.
When $\theta \neq 0$, neither of the two radial trap axes aligns with the propagation direction of the Raman laser, and the momentum kick applied to the ion excites the ion motion in both radial trapping modes. Consequently, the separated wave packets generated by the first $\pi /2$ pulse overlap when the trapping periods of the two radial axes coincide, resulting in the beat envelope observed in the interference signal.

For signal (ii) shown in Fig.~\ref{fig:fig2}(c), the interference peaks show a uniform height and lack a beat envelope despite the presence of two radial modes with different trap frequencies. This result clearly indicates that one of the trap axes aligns with the propagation direction of the Raman laser.
The period of the interference peaks is consistent with the trap frequency of 1338~kHz, representing the higher-frequency mode of the two radial axes.
This observation implies that the trap axis in the tightly confined direction within the two radial trap axes aligns with the propagation direction of the Raman laser.

In Fig.~\ref{fig:fig2}(d), the blue diamonds indicate the angle $\theta$ derived by fitting the experimentally obtained interference signals using Eq.~\ref{eq:eq1}. The angle $\theta$ starts near -50 degrees on the lower side relative to the GND offset voltage at the coupling point. As the GND offset voltage approaches the coupling point, $\theta$ gradually increases, indicating rotation of the trap principal axis. At the center of the coupling point, as indicated by (ii) in Fig.~\ref{fig:fig2}(d), $\theta$ reached zero, signifying the alignment of the trap principal axis with the horizontal direction of the system. After passing through the coupling point, $\theta$ approaches approximately 50 degrees. The overall trend of the rotation of the trap principal axes is depicted above Fig.~\ref{fig:fig2}(b).
A limiting angle of $\pm 50$ degrees shown in Fig.~\ref{fig:fig2}(d) indicates that the angle between the loosely confined trap axis and vertical axis reaches approximately 50 degrees at both extremes. This angle is consistent with the angle of $\pm 54$ degrees (shown by the two horizontal dashed lines), which is the angle between the line connecting the two GND electrodes and the vertical axis of the trap system.

\section{\label{sec:level4}Simulation}
The trap frequencies are calculated using electromagnetic field simulation. The trajectory of a charged particle with the mass and charge of a $\rm ^{171}Yb^+$ ion is simulated at the center of the trap.
Once the trajectory is calculated, it was Fourier-transformed to derive the trap frequencies and determine the directions of the trap principal axes by identifying the principal directions of the ion motion.

\begin{figure}[tb]
	\centering
	\includegraphics[width=6cm]{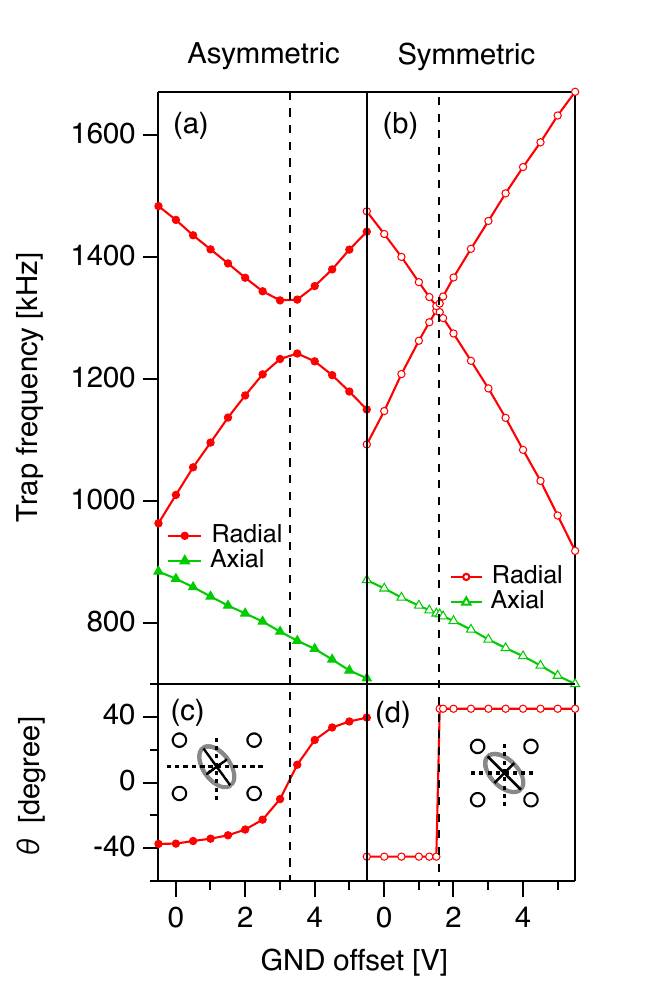}
	\onecolumngrid
	\caption{
	Calculated trap frequencies as a function of the GND offset voltage (a) in an asymmetric trap configuration, and (b) in a symmetric trap configuration. The angle of the trap principal axis to the vertical direction is calculated for the case of (c) an asymmetric trap configuration, and (d) a symmetric trap configuration. 
	}
	\label{fig:fig3}
\end{figure}

Figure~\ref{fig:fig3}(a) shows the GND offset voltage dependence of the trap frequencies obtained from the simulation of an asymmetric electrode configuration that mimics the actual trap configuration.
In this simulation, setting the RF amplitude and DC voltage for axial confinement identical to the experimental conditions results in a slight variance in the trap frequencies, which is attributed to imperfections in the electrode arrangement. To rectify this variance, we fine-tuned these parameters to accurately replicate the observed behavior of the trap frequencies in the simulation.
The closed circles and closed triangles in the figure show the trap frequencies in the radial and axial directions, respectively, as calculated from the simulation.
The measurement results shown in Fig.~\ref{fig:fig2}(a) and the simulated results shown in Fig.~\ref{fig:fig3}(a) agree closely in the trend of the trap frequency dependence on the GND offset voltage.
Furthermore, the radial biaxial coupling observed experimentally is reproduced in the simulation.
The frequency difference at the coupling point of the two axes in the asymmetric electrode configuration is calculated as $90~\rm kHz$.

However, no coupling between the radial confinements is observed in the simulation with the symmetric electrode configuration, as shown in Fig.~\ref{fig:fig3}(b). The open circles in Fig.~\ref{fig:fig3}(b) show the simulated trap frequencies in the two radial directions, and the open triangles represent the trap frequencies in the axial direction. The GND offset voltage of the coupling point is calculated to be $\rm 1.6~V$, deviating from the asymmetric trap configuration owing to variances in the trap electrode configuration. In contrast to the case of the asymmetric electrode layout, the two radial axes coincide perfectly, indicating that a circular potential can be formed in the radial plane.

We describe the angle $\theta$ of the trap principal axis relative to the vertical axis as calculated from the simulation.
The closed circles in Fig.\ref{fig:fig3}(c) and the open circles in Fig.\ref{fig:fig3}(d) show the values of the angle $\theta$ obtained from the simulation in the asymmetric and symmetric trap electrode configurations, respectively.
In the case of the asymmetric electrode layout, the angle rotates smoothly with the change in trap frequencies around the coupling point, as shown in Fig.~\ref{fig:fig3}(c).
Angle $\theta$ becomes zero precisely at the center of the coupling, as indicated by the vertical dashed line.
Consequently, at the coupling point, the loosely confined axis aligns with the vertical axis, whereas the tightly confined axis aligns with the horizontal axis.
On the lower-voltage side of the coupling point, the angle asymptotically approaches -40~degrees, whereas on the higher-voltage side, it asymptotically approaches 40~degrees.
In contrast, with a symmetrical arrangement of electrodes, the angle undergoes a stepwise change at the intersection of the two radial axes.
The angle remains constant at -40~degrees until reaching the coupling point, where it suddenly shifts to 40 degrees. This abrupt change occurs because the change in the GND offset voltage distorts the potential along the principal axis of the trap. As the GND offset voltage increases, the trap potential transitions from being elliptical at the lower voltage side, reaches the point of a perfectly circular condition, and then become elliptical again, aligning perpendicularly to the original elliptical potential.

\section{\label{sec:level5}Conclusion}
In this study, we investigated the control mechanism of the trap frequencies and the orientation of the ion trap principal axes. The direction of the trap axes, which is crucial for ion motion control, is determined not only by the trap electrode structure but also by the experimental parameters chosen for the ion trap. In actual ion-trap experiments, a DC voltage is applied to the trap electrodes to intentionally distort the trap potential and fine-tune the trap frequencies for various purposes. However, this tuning method also results in rotation of the trap principal axes. Our study successfully determine the trap axis orientation using matter-wave interference analysis driven by a Raman transition, revealing complex peak structures indicative of the axis angles. Electromagnetic field simulations confirmed our experimental findings, verifying the rotational trap axis behavior and frequency changes. The findings of this study shed light on the control mechanism of trap frequencies and the orientation of the ion trap principal axes, offering valuable insights into ion traps for diverse applications in quantum science and technology. 

This work is supported by JST-Mirai Program Grant Number JPMJMI17A3, Co-creation Place Formation Support Program (JPMJPF2015), and JSPS KAKENHI Grant Number JP23K13046.



\end{document}